\documentclass[twocolumn,runningheads]{svjour2}
\smartqed  
\usepackage{graphicx}
\journalname{Astrophysics and Space Science}
\begin{document}

\title{High Energy Processes in Pulsar Wind Nebulae
}


\author{W. Bednarek  
}


\institute{W. Bednarek \at
	      Department of Experimental Physics, University of \L \'od\'z, 90-236 \L \'od\'z, 
	      ul. Pomorska 149/153, Poland \\
              Tel.: +48-42-6355645 \\
              \email{bednar@fizwe4.fic.uni.lodz.pl }           
}

\date{Received: date / Accepted: date}

\maketitle

\begin{abstract}
Young pulsars produce relativistic winds which interact with 
matter ejected during the supernova explosion and the surrounding interstellar gas. 
Particles are accelerated to very high energies somewhere in the pulsar 
winds or at the shocks produced in collisions of the winds with the 
surrounding medium. As a result of interactions of relativistic leptons with the magnetic field and low energy radiation (of synchrotron origin, thermal, or microwave background), the non-thermal radiation is produced with the lowest possible energies up to $\sim$100 TeV. The high energy (TeV) $\gamma$-ray emission has been originally observed from the Crab Nebula and recently from several other objects. Recent observations by the HESS Cherenkov telescopes allow to study for the first time morphology of the sources of high energy emission, showing unexpected spectral features. They might be also interpreted as due to acceleration of hadrons.  
However,  theory of particle acceleration in the PWNe and models 
for production of radiation are still at their early stage of development since it becomes clear that realistic modeling of these objects should include their time evolution and three-dimensional geometry. In this paper we concentrate on the attempts to create a model for the high energy processes inside the PWNe which includes existence not only relativistic leptons but also hadrons inside the nebula.
Such  model should also take into account evolution of the nebula in time. Possible high energy expectations  based on such a model are discussed in the context of new observations.

\end{abstract}

\section{Introduction}

The non-thermal nebulae around young energetic pulsars (Pulsar Wind Nebulae - PWNe) have been suspected to accelerate leptons to sufficiently large energies allowing production of TeV
$\gamma$-rays (e.g. Gould~1965, Rieke \& Weekes~1969, Grindlay \& Hoffman~1971, Stepanian~1980). In fact, the nebula around the Crab pulsar was originally reported by the Whipple group as a first TeV $\gamma$-ray source (Weekes et al.~1989).  
Later, detections of  TeV $\gamma$-rays from nebulae around other pulsars have also been  claimed 
(e.g. around PSR 1706-44 - Kifune et al.~1995, Chadwick et al.~1998, Vela pulsar - Yoshikoshi et al.~1997, or PSR 1509-58 (MSH 15-52) - Sako et al.~2000),  but some of  the reported emission features have not been confirmed by recent more sensitive observations (see e.g. the results of HESS Collab. concerning PSR 1706-44 - Aharonian et al.~2005a, or Vela pulsar - Aharonian et al.~2006a).  Recent advances in the study of the pulsar wind nebulae obtained mainly by observations in the lower energy range (from radio to X-ray emission) are reviewed by Gaensler \& Slane~(2006).

Early detections of the TeV $\gamma$-ray emission from the PWNe have been usually interpreted in the so called synchrotron self-Compton model (SSC model) according to which
relativistic leptons inside the nebula produce soft synchrotron photons. These photons are next
up-scattered by the same leptons to the $\gamma$-ray energies as a result of inverse Compton process (IC).  Also soft photons  of other nature, microwave background radiation (MBR), infrared or optical
background present inside the PNW, are up-scattered by these leptons to $\gamma$-ray energies. The semi-phenomenological model of this type
has been elaborated by de Jager \& Harding~(1992) and successfully applied to the best studied TeV $\gamma$-ray source around the Crab pulsar. These authors apply this model assuming the distribution of the magnetic field inside the pulsar wind nebula obtained by Kennel \& Coroniti~(1984).
Using the  known spatial distribution of the synchrotron emission inside the nebula they get the spectra and distribution of relativistic leptons inside the nebula. Next, from the obtained
 distribution of leptons and soft radiation inside the nebula, they calculate the TeV $\gamma$-ray emission (spectra and spatial distribution). Based on such procedure, de Jager \& Harding predict that the higher energy TeV $\gamma$-ray emission should originate closer to the pulsar. The model has been up-dated by Atoyan \& Aharonian~(1996) by including other possible soft radiation targets inside the nebula and more recently confronted with the observations of the Crab Nebula by Hillas et al.~(1998). For the application of SSC model to nebulae around other pulsars see e.g. du Plessis et al.~(1995) - PSR 1509-58, or de Jager et al.~(1996) - Vela pulsar. 

The classical model of de Jager \& Harding~(1992) applies the multiwavelength observations of the Crab Nebula predicting the TeV $\gamma$-ray spectra at the present time. It does not consider the evolution of the nebula in time. 
The first simple time dependent leptonic model for the PWNe has been discussed by Aharonian, Atoyan \& Kifune~(1997) and applied to the nebula around PSR 1706-44. The authors analyze the evolution of the equilibrium energy spectrum of leptons injected at a constant rate, with a fixed spectrum (independent of time) into the region with a constant magnetic field. They come to an interesting conclusion that PWNe with relatively low magnetic field should become strong sources of $\gamma$-ray emission. Therefore, the efficiency of energy conversion from the rotating pulsar to $\gamma$-rays should increase with the age of the nebula.

Observation of the TeV $\gamma$-ray emission up to $\sim$80 TeV by the HEGRA Collab.
(Aharonian et al.~2004) indicates the existence of particles with energies up to $\sim$10$^{15}$ eV inside the Crab Nebula. Acceleration of leptons to such energies  requires special conditions in the acceleration region (low synchrotron energy losses, very efficient acceleration mechanism). Note however, that recently the HESS Collab. (Aharonian et al.~2006b) reported the $\gamma$-ray spectrum from the Crab Nebula showing an exponential cut-off at energy
$\sim$14 TeV. If real such discrepancy might suggest the presence of an additional component in the Crab Nebula spectrum above several TeV, which could be interpreted as the 
contribution from relativistic hadrons inside the nebula (see e.g. Bednarek \& Bartosik~2003, BB03). In fact, the existence of relativistic hadrons inside the PWNe has been considered in the past by e.g., Cheng et al.~(1990), Atoyan \& Aharonian~(1996), and Bednarek \& Protheroe~(1997). These hadrons interact with the matter of the supernova remnant and surrounding medium producing pions which decay to $\gamma$-rays and neutrinos.
Possible contribution of $\gamma$-rays from hadronic processes to the observed flux from the Crab Nebula has been discussed in the above mentioned papers. 
For example, Atoyan \& Aharonian~(1996) consider interaction of relativistic leptons and hadrons with the matter inside the Crab nebula and conclude that bremsstrahlung and $\pi^{\rm o}$ processes might give some interesting contribution to the observed spectrum provided that the effective density of matter inside the nebula is by one order of magnitude larger than the observed average density of matter. In this paper we discuss a more complete hadronic-leptonic model which has been recently proposed independently by Bednarek \& Bartosik~(2003) and  Amato, Guetta \& Blasi~(2003). Let us first review some new observational results in which context this model will be considered.

\section{New observations of the PWNe}

Since the aim of this paper is to review the expected multi-messenger high energy signatures for acceleration of particles in the PWNe, here we only mention the most interesting (in our opinion) new observational results obtained mainly by the HESS Collaboration in the TeV
$\gamma$-ray energy range.  For a more complete review of the HESS results see e.g., de Jager~(2006), Gallant~(2006), or Aharonian~(2006). 

The great advantage of the HESS Cherenkov telescope system is the ability of morphological studies of TeV $\gamma$-ray emission. It has appeared that TeV $\gamma$-ray emission from the middle-aged PWNe is extended and comes from the region up to several parsecs, e.g. MSH 15-52 (Aharonian et al.~2005b).
Moreover, the emission is often asymmetric which is probably the result of interaction of the supernova with the surrounding medium, e.g. PWNe associated with PSR B1823-13 (Aharonian et al.~2005c),  Kookaburra complex  (Aharonian et al.~2006c) or around the Vela pulsar (Aharonian et al.~2006a).  Such asymmetries are difficult to explain by the movement of the pulsars through the nebulae (e.g. the observed offset of the Vela pulsar by $\sim$0.5 degree would require its velocity of $\sim$250 km s$^{-1}$, while the measured value is only 65 km s$^{-1}$,  Caraveo et al.~2001).
Another interesting result of the HESS Collab. is the discovery of $\gamma$-ray emission between  550 GeV up to 65 TeV from the region of the Vela pulsar with unexpected very flat spectrum (index $\sim$1.45) and the cut-off at $\sim$14 TeV (Aharonian et al.~2006a). 
This extended TeV emission offset from the location of the Vela pulsar has been also confirmed the CANGAROO Collab. (Enomoto et al.~2006). 
In fact, all other PWNe have the spectral indices in the range $\sim$2.2-2.4, i.e. 
more similar to that observed for the Crab Nebula. The TeV $\gamma$-ray spectral features 
observed from the Vela nebula seem to be more consistent with their hadronic origin (e.g. Bednarek \& Bartosik~2003, Horns et al.~2006).

Since in the last few years no essential development in the  modeling of the high energy emission from the PWNe in the framework of the pure leptonic models has been obtained, we concentrate here on the recently considered hadronic-leptonic (hybrid) models which try to include not only radiation processes due to the injected leptons but also processes due to injection of relativistic hadrons into the nebula. These recent works were stimulated mainly
by the studies of the mechanism of possible acceleration of leptons as a result of their
interaction with the Alfven waves generated by coherently gyrating heavy nuclei in the PWNe
(for review see Arons~1998). Such hybrid leptonic-hadronic models also predict additional high energy messengers (neutrinos, neutrons or charged particles contributing to the observed cosmic rays at the Earth), whose observation will certainly put new constraints on the high energy processes occurring in the PWNe.  

\section{Hadronic-leptonic (hybrid) model}

Possible acceleration of heavy nuclei, extracted from the surface of the neutron star, by the electric field induced in the outer gaps of the inner pulsar magnetosphere was suggested already by Cheng et al.~(1986). However,  these nuclei should partially photo-disintegrate when passing through the non-thermal radiation of the outer gaps.  Neutrons from their disintegration are injected into expanding nebula surrounding the energetic pulsar. The decay products of neutrons, i.e. protons,
partially captured inside the nebula, interact with the matter and contribute to the $\gamma$-ray emission. Such model was suggested for the $\gamma$-ray and neutrino production from the Crab Nebula  (Bednarek \& Protheroe~1997). 

The nuclei can be additionally accelerated by the pulsar wind carrying significant amount of energy lost by the pulsar. They can  energetically dominate the relativistic pulsar wind
as proposed by Hoshino et al.~(1992) and Gallant \& Arons~(1994).  Based on the observations of the wisps in the Crab Nebula, these authors conclude that  the nuclei are accelerated  somewhere inside the pulsar wind zone (without specifying the details of the unknown acceleration process).  After passing the pulsar wind shock, the nuclei generate Alfven waves which energy is next 
absorbed by leptons. As a result, monoenergetic nuclei and leptons with a power law spectrum are injected into the nebula. The radiation model for the PWNe based on the above mentioned acceleration mechanism of leptons has been independently developed by Amato, Guetta \& Blasi~(2003) and Bednarek \& Bartosik~(2003). These models try to
include (for the first time self-consistently)  relevant leptonic and hadronic radiation processes inside the nebula taking into account also the time evolution of the nebula. We describe the main features of such a model and obtain results based on the approach presented in Bednarek \& Bartosik. 
Note, that although  both models differ in some details, they give not very different results  in the case of the Crab Nebula the concerning contribution of hadronic processes to the highest energy part of the $\gamma$-ray spectrum and possible detection of the pulsar wind nebulae by the future neutrino telescopes. Such hybrid hadronic-leptonic  model has been also considered as possible explanation of the $\gamma$-ray emission from the recently discovered $\gamma$-ray nebula associated with the Vela pulsar (Horns et al.~2006). 

\subsection{A simple model for expansion of the nebula}

The time evolution of a supernova remnant under the influence of an energetic pulsar is described according to the general picture proposed by Ostriker \& Gunn~(1971) and Rees \& Gunn (1974). Let us denote the initial expansion velocity of the bulk matter in supernova envelope by $V_{\rm 0,SN}$ and its initial mass by $M_{\rm 0,SN}$.
The  expansion velocity can increase due to additional supply of energy to the nebula by the pulsar. It can also decrease due to accumulation of  the  surrounding matter.
These processes are taken into account in order to determine the radius of the nebula at a specific time, t, by using the energy conservation,
\begin{eqnarray}
{{M_{\rm SN}(t)V_{\rm SN}^2(t)}\over{2}} = {{M_{\rm 0,SN}V_{\rm 0,SN}^2}\over{2}}
+ \int^t_0L_{\rm pul}(t')dt'
\label{eq1}
\end{eqnarray}
\noindent
where 
\begin{eqnarray}
L_{\rm pul}(t) = B_{\rm s}^2 R_{\rm s}^6 \Omega^4/6c^3\approx 
3\times 10^{43}B_{12}^2P_{\rm ms}^{-4}~~{\rm erg~s}^{-1}, 
\label{eq2}
\end{eqnarray}
\noindent
is the pulsar energy loss, $\Omega = 2\pi/P$, and $P = 10^{-3}P_{\rm ms}$ s
changes with time according to 
$P^2_{\rm ms}(t) = P_{\rm 0,ms}^2 + 2\times 10^{-9}tB_{12}^2$, 
$P_{\rm 0,ms}$ is the initial period of the pulsar, and $B = 10^{12}B_{12}$ G is 
the surface magnetic field of the pulsar.

\noindent
The nebula increases its mass according to
\begin{eqnarray}
M_{\rm SN}(t) = M_{\rm 0,SN} + {{4}\over{3}}\pi \rho_{\rm sur} R_{\rm Neb}^3(t),
\label{eq4}
\end{eqnarray}
\noindent
where $\rho_{\rm sur}$ is the density of surrounding medium and $R_{\rm Neb}$ 
is the radius of the expanding envelope at the time {\it t},
\begin{eqnarray}
R_{\rm Neb} = \int_0^t V_{\rm SN}(t')dt'.
\label{eq5}
\end{eqnarray}
The average density of matter inside the nebula is
\begin{eqnarray}
\rho_{\rm Neb} = 3M_{\rm SN}(t)/4\pi R_{\rm Neb}^3(t).
\label{eq6a}
\end{eqnarray}
\noindent
The above set of equations is solved numerically in order to obtain the basic parameters of the expanding PWN.

Also the location of the inner shock inside the PWN can be estimated by  balancing 
the pressure of the expanding nebula with the pressure of the pulsar wind (Rees \& Gunn~1974),
\begin{eqnarray}
{{L_{\rm pul}(t)}\over{4\pi R_{\rm sh}^2c}}\approx 
{{\int_0^t\sigma L_{\rm pul}(t')dt'}\over{{{4}\over{3}}\pi R_{\rm Neb}^3}},
\label{eq6}
\end{eqnarray}
\noindent
where $\sigma$ is the ratio of the magnetic energy flux to the total energy 
flux lost by the pulsar at the location of the pulsar wind shock at radius 
$R_{\rm sh}$. 
Then, the magnetic field strength at the shock region can be estimated from
\begin{eqnarray}
B_{\rm sh} = \sqrt{\sigma}B_{\rm pul}\left({{R_{\rm pul}}\over{R_{\rm lc}}}\right)^3
{{R_{\rm lc}}\over{R_{\rm sh}}},
\label{eq10}
\end{eqnarray}
\noindent
where $R_{\rm pul}$ and $B_{\rm pul}$ are the radius and the surface magnetic field 
of the pulsar, and $R_{\rm lc}$ is the light cylinder radius.

The evolution of basic parameters characterizing the pulsar wind nebula, for some initial parameters of the supernova  explosion and the pulsar, are shown on Fig.~1 in BB03.

\subsection{Model for acceleration of leptons and hadrons}

As already noted above we are interested in the scenario in which rotating magnetospheres of neutron stars can accelerate not only leptons but also heavy nuclei, extracted from
positively charged polar cap regions. In fact, different aspects of the high energy
phenomena around pulsars, such as the change in the drift direction of the radio
sub-pulses (Gil et al.~2003), the existence of morphological features inside the Crab 
Nebula (so called radio wisps), and the appearance of extremely energetic leptons inside it 
(Gallant \& Arons 1994), can be naturally explained by the presence of heavy nuclei.
Arons and collaborators (e.g. see Arons~1998) postulate that the Lorentz factors of iron nuclei 
accelerated somewhere in the inner magnetosphere and/or the Crab pulsar 
wind zone should be,
\begin{eqnarray}
\gamma_{Fe}\approx \eta Ze\Phi_{\rm open}/m_{\rm Fe}c^2\approx
8\times 10^9 \eta B_{12} P_{\rm ms}^{-2},
\label{eq1}
\end{eqnarray}
\noindent
where $m_{\rm Fe}$ and $Ze$ are the mass and charge of the iron nuclei,
$c$ is the velocity of light, and $\Phi_{\rm open} = \sqrt{L_{\rm pul}/c}$
is the total electric potential drop across the open magnetosphere,
and $\eta$ is the acceleration factor determining the Lorentz factor of 
nuclei in respect to the maximum one allowed by the pulsar electrodynamics.
Following Arons and collaborators, 
the authors assumed: (1) $\eta$ is not very far from unity, the value $\eta = 0.5$ is adopted; 
(2) iron nuclei take most of the spin down power of the pulsar, 
$L_{\rm Fe} = \chi L_{\rm rot}$, where $\chi = 0.95$. 
Unfortunately, this values are not predicted at present by 
any model of the ion acceleration in the pulsar wind and can only be constrained  
by the high energy observations of the PWNe.
The iron nuclei are extracted from the neutron star surface and accelerated
during the pulsar radio phase when the efficient leptonic cascades heat
the polar cup region. They are farther accelerated in the pulsar wind zone due to the linear increase of the Lorentz factor of the outflowing plasma (as more recently discussed by
Contopoulos \& Kazanas~2002) or due to reconnection of oppositely directed magnetic fields in the wind (Michel~1982, Coroniti~1990, Lyubarsky \& Kirk~2001). 
The quasi-monoenergetic heavy nuclei, after crossing the pulsar wind shock, generate Alfven waves in the down-stream region, which energy is resonantly transfered to leptons present in the wind (Hoshino et al.~1992). 
As a result, leptons are accelerated to high energies with  a power law spectrum with the
spectral index $\delta_1\approx 2$ between $E_{\rm 1} = \gamma_{\rm Fe}m_{\rm e}c^2$ 
and $E_{\rm 2}\approx \gamma_{\rm Fe} A m_{\rm p}c^2/Z$ (see Gallant \& Arons~1994), 
where $m_{\rm e}$ and $m_{\rm p}$ are the electron and proton mass, respectively.
The spectrum is normalized in such a way as to get the conversion efficiency of energy from the iron nuclei 
to the leptons equal to  $\xi$. Note that the radiation from  
leptons depends on the product of the energy conversion from the pulsar wind
to nuclei, $\chi$, and the acceleration efficiency of positrons by these ions, $\xi$.
Therefore, decreasing the first coefficient and increasing the second, one
obtains the same level of radiation from positrons but a lower level of gamma-ray
flux from hadronic interactions of ions with the matter inside the nebula.
Since the dependence of $\chi \cdot \xi$ on time is not predicted by any 
theoretical model in the original paper by Bednarek \& Bartosik~(BB03), this value was kept constant during the evolution of the nebula. 
Relativistic particles accelerated by the mechanism discussed above are captured 
inside the pulsar wind nebula losing energy on different processes. 
However, nuclei injected into the nebula at a specific time $t_{\rm inj}$,
escape from it at the time $t_{\rm esc}$, if their diffusion distance in the turbulent magnetic
field of the nebula, $R_{\rm diff}$, is equal to the dimension of the nebula,
$R_{\rm Neb}$, at the time $t_{\rm esc}$. The diffusion distance in the magnetic field
of the nebula is obtained by integration (Bednarek \& Protheroe~2002),
\begin{eqnarray}
R_{\rm diff} = \int^{t_{\rm esc}}_{t_{\rm inj}} \sqrt{{{3D}\over{2t'}}}dt',
\end{eqnarray}
\noindent
where the diffusion coefficient is taken to be $D = R_{\rm L}c/3$, and
$R_{\rm L}$ is the Larmor radius of nuclei depending on the distance from the center of the
nebula. $R_{\rm diff}$ and $t_{\rm esc}$ are calculated numerically.
The adiabatic losses of nuclei during their propagation inside the nebula are also included. Due to these losses and interactions of nuclei with the matter the energy
of a nucleus at a specific time is
\begin{eqnarray}
E(t) = E(t_{\rm inj}) {{t_{\rm inj} + t}\over{2t_{\rm inj}K^\tau}},
\end{eqnarray}
\noindent
where $E(t_{\rm inj})$ is its energy at
the time $t_{\rm inj}$, K is the inelasticity coefficient for collisions
of nuclei with matter,
and $\tau$ is the optical depth calculated from the known density of
matter inside the nebula. The details of these calculations are given in sections 4.2 and 4.3 in Bednarek \& Protheroe~(2002). All the above mentioned effects have been taken into account when considering leptons and hadrons inside the nebula.

\section{High energy radiation from the PWNe in the hybrid model}

Having defined the model of the nebula expansion (which gives the basic parameters of the nebula), and also the injection spectra of particles as a function of time after the supernova explosion (the age of the nebula), the equilibrium spectra of leptons and hadrons inside the nebula can be obtained taking into account different energy loss processes. 
Leptons injected into the medium of the expanding supernova 
remnant suffer energy losses mainly on radiation processes, bremsstrahlung, synchrotron, 
and the inverse Compton, and due to the expansion of the nebula. 
The rate of their energy losses can be described by 
\begin{eqnarray}
-{{dE}\over{dt}} = (\alpha_1 + \alpha_2)E + (\beta_1 + \beta_2) E^2~~~{\rm GeV~s^{-1}},
\label{eq22}
\end{eqnarray}
\noindent
where $\alpha_1$ and $\alpha_2$ describe the bremsstrahlung and adiabatic losses; $\beta_1$ and $\beta_2$ the synchrotron and ICS energy losses. The coefficients, $\alpha_1, \alpha_2, \beta_1$, and $\beta_2$, depend on time 
in a complicated way due to the changing conditions in the expanding nebula (magnetic
field, density of matter and radiation). Therefore, the above formula can not be integrated 
analytically at an arbitrary time after supernova explosion. 
In order to determine energies of leptons, $E$, inside the 
nebula at a specific time $t_{\rm obs}$,
which have been injected with energies $E_{\rm o}$ at an earlier time $t$, 
we use the numerical approach. 
However, knowing the parameters of the nebula determined at a given time $t$,  the 
evolution of the equilibrium spectrum of leptons during  the time step $\Delta t$ is determined analytically. Next, the conditions inside the
nebula are changed to values which are obtained from the expansion model of the nebula
at time $t+\Delta t$.
The equilibrium spectrum of leptons at, $t_{\rm obs}$, is then obtained  
by summing over the spectra injected at specific time and over all time steps
up to the present observed time $t_{\rm obs}$, 
\begin{eqnarray}
{{dN(t_{\rm obs})}\over{dE}} = \sum_{t=0}^{t=t_{\rm obs}} J(t')
 {{dN}\over{dE_{\rm o}}dt}dt,
\label{eq24}
\end{eqnarray}
\noindent
where $dN/dE_{\rm o}dt$ is the injection spectrum of leptons at time t, 
$t' = t_{\rm obs} - t$, and the jacobian $J(t') = E_{\rm o}/E(t')$ 
describes the change of energy of lepton during the period $\Delta t$. It
is calculated analytically by solving Eq.~11.
The example equilibrium spectra of leptons inside the nebula at the specific time after explosion of supernova are given in Fig.~3 in BB03. 
 
In order to calculate the equilibrium spectra of different types of nuclei inside the nebula at its given age, we take into account the energy losses on the interaction of nuclei with the matter of the nebula, the adiabatic energy losses, and the escape of nuclei from the nebula due to diffusion. The example equilibrium spectra of nuclei inside the nebula at specific age are shown in Fig.~2 in Bednarek \& Bartosik~(2003). 

\subsection{Gamma-rays}

Having obtained the equilibrium spectra of relativistic leptons and hadrons
as a function of time after supernova explosion we can calculate the energy 
spectra of photons produced inside the nebula by these particles in different radiation processes. In general, such spectra are characterized by three main components: two due to leptons (synchrotron, IC) and one due to hadrons
($\pi^{\rm o}$ decay). Depending on the parameters of the pulsar, the supernova, the surrounding medium and on the age of the nebula, different components may dominate.
For the PWNe in a very early stage of development (less than $\sim$100 yrs), $\gamma$-rays from the interaction of hadrons with matter  may dominate.
For young nebulae (at the age of $\sim$10$^3$ yrs, the Crab type), the power of the synchrotron spectrum dominates over that of  the IC $\gamma$-ray spectrum. In the case of the middle age nebulae (at the age $\sim$10$^4$ yrs, the Vela type), the power in synchrotron bump may become comparable to the power in the TeV $\gamma$-ray bump (produced mainly due to the IC scattering of the MBR). 

\begin{figure}
\vskip 5.5truecm
\includegraphics{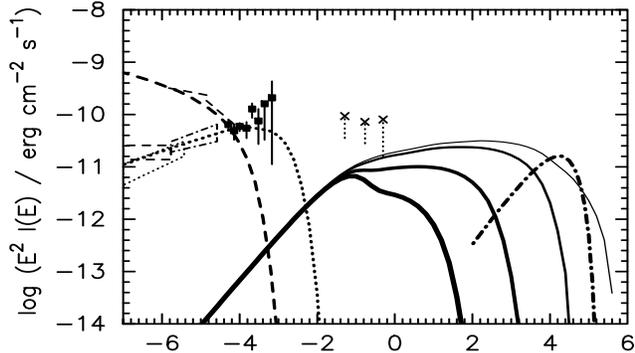}
\caption{The $\gamma$-ray spectrum from Vela nebula modeled with hybrid
hadronic-leptonic model (Bednarek \& Bartosik~2003). $\gamma$-rays from $\pi^{\rm o}$ 
decay created in hadronic collisions (dot-dashed curve) and from ICS of leptons (full curves).
The acceleration mechanism of leptons operates through all the lifetime of the pulsar
(the thinnest full curve), or only during the first 5000 yrs, 3000 yrs, and 2000 yrs (the thickest full curve). The acceleration of hadrons is continuous throughout the age of the pulsar $\sim$11000 yrs.
The dashed and dotted curves show the synchrotron emission from the extended and compact nebulae.}
\label{fig3}
\end{figure}
For reasonable initial parameters, the hybrid model predicts also the appearance of the small bump in the multiwavelength spectrum of the Crab nebula at the highest energies, see Bednarek \& Bartosik~(2003, 2005) or Amato, Guetta \& Blasi~(2003). In fact, recent measurements of the cutoff in the Crab Nebula spectrum at $\sim$14 TeV (Aharonian et al.~2006a), combined with the previous measurements of the spectrum up to $\sim$80 TeV (well described by a simple power law, Aharonian et al.~2004), might be interpreted as the evidence of such an additional component at $\sim$100 TeV. 

Based on the hybrid model, Bednarek \& Bartosik (BB03) calculated expected emission from the nebula around the Vela pulsar predicting significant contribution to the $\gamma$-ray  flux
from the decay of $\pi^{\rm o}$. However, in the original calculations it has been assumed that relativistic leptons are injected into the nebula as a result of resonant interactions with nuclei during the whole lifetime of the pulsar.
Recent results obtained by the HESS Collab. (Aharonian et al.~2006a) show that it may not be the case. As we noted above, the observed TeV $\gamma$-ray spectrum from the Vela Nebula is flatter than $E^{-2}$,  and very nicely resembles the pure hadronic component (see Fig.~7 in BB03). 
Fig.~1 shows  the $\gamma$-ray spectra obtained in terms of the hybrid model which are re-calculated applying the modified assumption about the acceleration process of leptons by hadrons inside the nebula. In the present calculations, leptons are accelerated only during the first 2000, 3000, and 5000 yrs after the pulsar birth. Leptons accelerated in the past lose efficiently their energy on radiative and adiabatic processes. Therefore, the equilibrium spectrum of leptons inside the nebula shifts to lower energies with increasing age of the nebula due to the lack of freshly injected leptons at the present time. As a result, the IC spectrum obtained in such a modified model cuts-off at lower energies and the flat $\gamma$-ray emission above $\sim$1 TeV due to the decay of $\pi^{\rm o}$ can be clearly observed. 
Such model predicts the appearance of a steep component in the TeV $\gamma$-ray spectrum below $\sim$1 TeV which might be used to estimate the duration of the acceleration process of leptons.

Due to the proximity of the pulsar and diffusion of leptons from the acceleration site close to the pulsar wind shock, it has been predicted that the TeV $\gamma$-ray emission from the Vela PWN should be extended with characteristic dimension of the order of $\sim$5 pc (Bednarek \& Bartosik~2005). This is consistent with the HESS observations. Note, that the Vela nebula is significantly displaced from the location of the Vela pulsar which might be the result of unequal pressure on the nebula due to its interaction with the inhomogeneous medium as earlier suggested by Blondin, Chavalier \& Frierson~(2001).   

Note, however that the morphological studies of the TeV $\gamma$-ray emission features of the pulsar wind nebula G18.0-0.7, close to the pulsar PSR J1825-137, show clear softening of the spectrum with increasing  distance from the pulsar (Aharonian et al.~2006d). Such feature is better explained by the leptonic model in which leptons lose radiatively and adiabatically energy when diffusing outside the pulsar. 

\subsection{Neutrinos and Neutrons}

Provided that the hypothesis on hadron acceleration in some PWNe is correct and assuming 
that significant amount of observed $\gamma$-ray flux is due to hadrons, it is possible to estimate the expected neutrino flux and the event rates in the presently planned and constructed 1 km$^2$ neutrino detectors (IceCube, KM$^3$NET). Such simple estimate, based on the observed TeV $\gamma$-ray flux, has been performed by Amato \& Guetta~(2003) for a few PWNe. For example these authors argue that $\sim$12 neutrino events per km$^3$ per year should be detected from the Crab Nebula.
More detailed calculations according to the hybrid model of Amato, Guetta \& Blasi~(2003)
predicts $\sim$5-13 neutrino events per yr per km$^3$ from the Crab Nebula, provided that hadron  Lorentz factors reach $\sim$10$^4$-10$^7$. The event rates estimated for the Crab and Vela pulsars, based on a similar model by 
Bednarek~(2003), are on the level of $\sim$1 neutrino event within 1 km$^2$ detector during 1 year, i.e at the level of the atmospheric neutrino background (ANB). However, when these nebulae were younger, then the event rates were significantly larger, decreasing with the age of the nebula, $t_{\rm neb}$, proportionally to $\sim t_{\rm neb}^{-2}$ (see Fig.~1 in Bednarek~2003).
Based on the observed fluxes of the TeV $\gamma$-rays from the PWNe by the HESS Collab., Kappes et al.~(2006) predicts on average $1.4-4.1$ neutrino events per year above $>$5 TeV from the direction of the Vela Nebula in the KM$^3$NET detector, whereas the estimated ANB is of the level of   $\sim$1 neutrino event.
   
The hybrid model for $\gamma$-ray production in the PWNe postulates also that from relatively close sources, the $\gamma$-rays and neutrinos might be accompanied by neutrons. In fact, $\gamma$-ray emission up to 
$\sim$65 TeV from the direction of the Vela Nebula, requires the acceleration of hadrons to the Lorentz factors of the order of $\sim$10$^6$. In the past, when  the Vela pulsar was at the present age of the Crab pulsar, nuclei might be accelerated to the Lorentz factors of the order of magnitude larger, i.e. $\sim$10$^7$. The mean free path for  neutrons extracted from nuclei with such Lorentz factors (or produced in hadronic interactions of protons) is already $\sim$100 pc. So then, significant number of neutrons from the Vela pulsar might also reach the Earth about $10^4$ yrs ago, from the distance of $\sim 300$ pc (Caraveo et al.~2001).    

\subsection{Contribution to Cosmic rays?}

Accepting the hybrid model and keeping in mind  that PWNe are at present the best established TeV $\gamma$-ray sources, we should expect  that particles escaping from the PWNe might contribute to the cosmic rays (CR) in the Galaxy. In fact, pulsars have been suspected since their discovery as main sources of cosmic rays. Ostriker \& Gunn~(1969) and Karaku\l a, Osborne \& Wdowczyk~(1974) postulated acceleration of CRs above the knee region in the pulsar wind zones by the large amplitude electromagnetic waves generated by rotating
neutron stars. The contribution of particles accelerated by pulsars to the observed cosmic ray spectrum has was later discussed by Cheng \& Chi~(1996), Bednarek \& Protheroe~(2002), Giller \& Lipski~(2002). Cheng \& Chi~(1996) propose that nuclei injected from the pulsar magnetospheres during the pulsar glitches can contribute to the knee region in the CR spectrum. Giller \& Lipski~(2002) derive the
initial parameters of the pulsar population inside the Galaxy required to explain the
observed shape of the CR spectrum and its intensity up to the highest energies. Bednarek \& Protheroe~(2002) estimate the contribution of heavy nuclei accelerated in the pulsar outer gaps (Cheng, Ho \& Ruderman~1986) from the population of Galactic radio pulsars to the CRs above the knee region taking into account the propagation and escape conditions, adiabatic and collisional energy losses of the nuclei during their propagation in the pulsar wind nebulae. In fact,
a new component in the CR spectrum at energies above $\sim 10^{15}$ eV, supplied e.g. by pulsars, is required by the measurements of the mass composition, which suggests an increase of the average mass above the knee (e.g. Glasmacher et al.~1999, Ave et al.~2003).

Based on the hybrid model, Bednarek \& Bartosik (2004) calculated the spectra of nuclei which escape from the PWNe around the pulsars with assumed initial parameters. Applying a few different models for the initial parameters of the pulsars formed inside the Galaxy, concerning the distribution of the surface magnetic fields of the new born neutron stars (derived by Narayan~1987) and the initial pulsar periods (derived by Lorimer et al.~1993), the contribution of the pulsar population to the CRs in  the Galaxy has been estimated. It is concluded that model B for the pulsar population presented in Lorimer at al.~(1993) is able to describe satisfactory the shape of the CR spectrum and the features of the mass composition in the energy range between the knee and the ankle (see Figs. 3b and 4 in Bednarek \& Bartosik~2004).
The fine features in the CR spectrum (the knee region)  might also appear due to the presence of the nearby pulsars (e.g. Erlykin \& Wolfendale~2004). However, such pulsars would create an anisotropy above the present observational limit (Bhadra~2006).

Also extremely high energy CRs, i.e. above the ankle at $\sim 10^{18}$ eV, might be accelerated in the wind regions of pulsars with super strong surface magnetic fields, so called magnetars. Blasi, Epstein \& Olinto~(2000), propose that the highest energy CRs are accelerated locally in our Galaxy by the pulsars whose initial periods are shorter than $\sim$10 ms  and whose surface magnetic fields are  $10^{12}-10^{14}$ G. If such pulsars inject heavy nuclei with energies $\sim 10^{20}$ eV, then their Larmor radii are smaller than the dimensions of the galactic halo, allowing their efficient trapping inside the
Galaxy. The authors argue that such nuclei can be accelerated to extremely high energies
after the moment when supernova envelope becomes transparent. However it is not
clear if they are not captured inside the supernova envelope  by the magnetic field of the PWN?  For example, from the modeling of the PWNe (BB03), it is possible to estimate the magnetic field inside the nebula, $\sim$10-100 G, and its dimension, $\sim 10^{15}-10^{16}$ cm, at $\sim$1 year after explosion.  For such parameters, the Larmor radius for iron nuclei are lower than the dimensions of the nebula allowing an efficient  trapping. Then, the adiabatic energy losses of nuclei due to fast expansion of the nebula become important. In a similar model, Arons~(2003) suggests that pulsars with extreme parameters produced in the whole Universe are responsible for the highest energy CRs. 
The winds of such pulsars disrupt the supernova envelopes allowing in this way acceleration of protons to extremely high energies (of the order of $10^{21-22}$ eV), and their escape from the nebula. The final shape of the CR spectrum at the highest energies in the universe depends on the rate of gravitational energy losses by the pulsar during its early stage after formation.

Note that the PWNe associated with such very young pulsars in the Galaxy should also produce significant fluxes of neutrinos which might be detectable within a few years after their birth, see e.g. Berezinsky \& Prilutsky~(1978), Protheroe, Bednarek \& Luo~(1998), Beall \& Bednarek~(2002). 

\section{Interaction of PWNe with massive clouds}

As reported in Sect.~2, the TeV sources are often substantially displaced from the positions of the associated middle aged pulsars. This is probably the effect of the interaction of a parent supernova shock wave with the surrounding inhomogeneous medium. This expectation seems to be very natural since young supernovae are usually close to the star forming regions containing  massive clouds. Particles  escaping from the PWNe can be partially captured by these high density clouds producing $\gamma$-rays in collisions with the cloud matter.
In fact, the TeV $\gamma$-ray emission from  one of the most massive compact stellar association, Cyg OB2, has been 
detected by HEGRA Collab. (Aharonian et al.~2002). Observation of other 14 young open clusters does not give positive detections due to the insufficient sensitivity (Aharonian et al.~2006e). Recently, extended TeV $\gamma$-ray emission, coincident with the location of giant molecular clouds, has been also observed from the direction of the Galactic Center 
(Aharonian et al.~2006f). Such extended $\gamma$-ray sources can be also caused by the appearance of nearby energetic pulsars. If such pulsars have been born typically 
$\sim$10$^4$ yrs ago with initial periods of the order of a few millisecond and the surface magnetic fields of the order of $\sim$10$^{13}$ G, then at present they should have periods close to $\sim$200 ms. Such radio pulsars are difficult  to discover in the high density regions. Hadrons accelerated during the lifetime of a pulsar can be partially captured in dense ($10^3-10^5$ cm$^{-3}$), magnetized 
($10^{-5}-10^{-3}$ G) molecular clouds, producing $\gamma$-rays for a relatively long time.
Note, that even hadrons with Lorentz factors up to $\sim$10$^9$ might be captured in the molecular clouds with dimensions of several parsecs since their Larmor radii are small enough.
Possible radiation effects from the interactions of the PWNe with high density regions have been considered in the context of the $\gamma$-ray emission from the Galactic Center (Bednarek~2002), and also proposed as a possible explanation of the unidentified TeV $\gamma$-ray source in the Cyg OB2 region (Bednarek~2003).  

\section{Conclusion}

Pulsars produce  a class of the best established TeV $\gamma$-ray sources in which particles are accelerated above $\sim$100 TeV.  Observations  with the HESS telescopes allow for the first time to study morphology  of these objects indicating that the interaction of parent supernova remnants with the surrounding medium strongly influences the evolution of the pulsar wind nebulae. These interaction effects have not been considered in the past leptonic models for the production of high energy radiation inside PWNe and strongly encourage  the efforts to work on a much more advanced, time dependent, three dimensional model for the PWNe. 
Moreover, new observations of the PWNe at TeV $\gamma$-rays seem to suggest that acceleration of hadrons to high energies inside the PWNe  should be also seriously considered. In fact, the unusual shape of the $\gamma$-ray spectrum from the nebula associated with the Vela pulsar (relatively narrow peak at several TeV) and possibly weak evidence of a discrepancy between observations of the Crab Nebula at the highest energies by HEGRA and HESS  can be naturally explained by the existence of an additional component in the TeV $\gamma$-ray spectra from these two objects which is due to contribution from relativistic hadrons. Since hadrons are weakly coupled to matter, they can diffuse from the
PWNe, interact with the surrounding  medium, and also contribute to the observed cosmic ray spectrum in the Galaxy. Therefore, PWNe should be considered as one of  the serious candidates responsible for  the bulk of the cosmic rays above the knee region, i.e. $>$10$^{15}$ eV in the Galaxy. Detection (or non-detection) of other neutral radiation from the PWNe, i.e. neutrinos and neutrons, should give the answer to this question. 

The aim of this paper was to stress the importance of such multi-messenger high energy observations of  processes occurring inside the PWNe.

\begin{acknowledgements}
I would like to thank M. Giller for reading the manuscript.
This research is supported by the Polish grant 1P03D01028.
\end{acknowledgements}



\end{document}